\documentstyle[12pt]{article}

\begin{document}
\baselineskip 23pt
\begin{center}
{\Large \bf The Two-dimensional Nonlinear \\
Burridge-Knopoff Model of Earthquakes}\\
\vskip1cm
{\bf Kwan-tai Leung} \\[2mm]
Institute of Physics, Academia Sinica, \\
Nankang, Taipei 11529 Taiwan, R.O.C. \\[8mm]
{\bf Judith M\"{u}ller}
and {\bf J\o rgen Vitting Andersen}
\footnote{Present address:
Laboratoire de Physique de la Mati\`{e}re Condens\'{e}e,
Universit\'{e}\ de Nice-Sophia Antipolis,
Parc Valrose, 06108 Nice Cedex 2, France}
\\[2mm]
Department of Physics, McGill University,\\
Rutherford Building, 3600 University Street,\\
Montr\'{e}al, Qu\'{e}bec, Canada H3A 2T8.\\[1cm]
\end{center}

\centerline{\bf Abstract}
We present a two-dimensional spring-block model of earthquakes with the full
nonlinear equations for the forces in terms of displacements. Correspondence
of our linearized version to earlier models reveals in them inherently
asymmetric elastic properties.  Our model generalizes those models and allows
an investigation of the effects of internal strains and vectorial forces. The
former is found to be relevant to critical properties such as that described
by the Gutenberg-Richter law, but the latter as well as nonlinearities up to
second order in displacements are irrelevant.

\vspace{1cm}

\newpage
In 1967, Burridge and Knopoff\cite{Burridge} introduced a one-dimensional
(1D) system of springs and blocks
to study the role of friction along a fault
in the propagation of an earthquake. Since then,
many other researchers investigated
similar dynamical models of many-body systems with
friction, ranging from propagation and rupture in earthquakes
[2-10] to the fracture of overlayers on a rough substrate\cite{Joergen}.

Among these developments,
a deterministic version of the
1D Burridge-Knopoff (BK) model was analyzed
by Carlson and Langer\cite{Langer}
and the same model but in a quasi-static limit
was studied by Nakanishi\cite{Nakanishi}.
A 2D quasi-static variant was first simulated
by Otsuka\cite{Otsuka} and later by Brown, Scholz and Rundle\cite{Brown},
who introduced a discrete version
that was formulated as a cellular automaton.
A similar model with
continuous local variables, generalizing
the model of Bak, Tang and Wiesenfeld\cite{Bak},
was studied by Olami, Feder and Christensen\cite{OFC} (OFC).
Contrary to previous models, the model is non-conservative and
derivable from a 2D BK model
(under certain limit, see below).
It produces robust critical behavior which depends on
the conservation level. OFC argued that such dependence explains
the variance of the exponent in the Gutenberg-Richter\cite{Gutenberg} law
observed in real earthquakes\cite{Pacheco}.

{\it Model.} ---
As before, our model consists of a 2D array of interconnected
blocks in contact with a rough surface.
Each block is also connected to a rigid driving plate by
a {\it leaf\/} spring whose spring constant is $K_L$.
The plate moves at constant, infinitesimal speed.
The coordinates for the attachment of the leaf springs on
the moving plate, labeled by
$(i,j)$ where $1\leq i,\,j \leq L$, form a square lattice with
lattice constants $a_1$, $a_2$ (hereafter the subscripts `1'
and `2' refer to properties in the $x$- and $y$-direction, respectively).
Driving the plate thereby
induces stress between the array and the plate.
Each block in the bulk is connected to
four nearest neighbors via {\it coil\/} springs
whose spring constants are $K_1$ and $K_2$
and unstretched lengths $l_1$ and $l_2$.
We restrict ourselves to the situation where
$a>l$, and the
displacements $x_{i,j}$, $y_{i,j}$ measured from $(i,j)$
fulfill $x_{i,j} \ll a_1$ and $y_{i,j} \ll a_2$,
so that Hooke's law applies.
When the net force on a block exceeds
a static frictional threshold $F_{th}$, the block slips instantaneously
to a new equilibrium position.

While this is an obvious extension of the
1D BK model\cite{Burridge} to two dimensions, equations from previous
models\cite{Otsuka,Brown,OFC} fail to describe it correctly.
The reason is that elements in those models have been
restricted to move in one direction only.
To our knowledge, this restriction has not been
justified beyond the reason of simplicity.
To address its relevance, we lift this limitation and start
with the proper nonlinear equations
for forces as functions of the displacements of blocks.
The net force $\vec{F}_{i,j}$ (which equals the friction due to
the rough surface) on the block at $(i,j)$ is a vector:
\begin{eqnarray}
\vec{F}_{i,j}& =& \vec{f}_L +
\vec{f}_{(i+1,j)-(i,j)} + \vec{f}_{(i-1,j)-(i,j)}
+  \vec{f}_{(i,j+1)-(i,j)} + \vec{f}_{(i,j-1)-(i,j)}
\label{totalforce}
\end{eqnarray}
where $\vec{f}_{(i\pm 1,j\pm 1)-(i,j)}$ is the force
exerted by a neighboring block
and $\vec{f}_L$ the loading force by the driving plate.
For example, we have
\begin{eqnarray}
f^x_{(i+1,j)-(i,j)}& =&   K_1 [ a_1 + x_{i+1,j} - x_{i,j}  -
  { ( a_1 + x_{i+1,j} - x_{i,j}  ) l_1 \over
\sqrt{ (a_1 + x_{i+1,j} - x_{i,j} )^2 + (y_{i+1,j} - y_{i,j})^2 }}], \\
f^y_{(i+1,j)-(i,j)}& =&   K_2 [ y_{i+1,j} - y_{i,j} -
{ ( y_{i+1,j} -  y_{i,j} ) l_2 \over
\sqrt{ (a_1 + x_{i+1,j} - x_{i,j} )^2 + (y_{i+1,j} - y_{i,j})^2 }}].
\label{components}
\end{eqnarray}
Then the $x$ component of the force in Eq.~(\ref{totalforce}) takes the form:
\begin{eqnarray}
F^x_{i,j}&=& f_L^x - K_1 [ 2 x_{i,j} - x_{i+1,j} - x_{i-1,j} +
\nonumber \\
& & { ( a_1+ x_{i+1,j} - x_{i,j}  ) l_1 \over \sqrt{
(a_1+x_{i+1,j}-x_{i,j})^2+(y_{i+1,j} - y_{i,j})^2 }} -
 { ( a_1-x_{i-1,j} + x_{i,j} ) l_1 \over \sqrt{
(a_1-x_{i-1,j}+x_{i,j})^2+(y_{i,j} - y_{i-1,j})^2 }}] -
\nonumber \\
& & K_2 [ 2x_{i,j} - x_{i,j+1} - x_{i,j-1} +
\nonumber \\
& & { ( x_{i,j+1} - x_{i,j} ) l_2 \over \sqrt{
(a_2+y_{i,j+1}-y_{i,j})^2 + (x_{i,j+1} - x_{i,j})^2 }} +
 {  ( x_{i,j-1} - x_{i,j} ) l_2 \over \sqrt{
(a_2-y_{i,j-1} + y_{i,j} )^2 + (x_{i,j} - x_{i,j-1})^2 }} ]
\label{xcomponent}
\end{eqnarray}
and, by symmetry, $F^y_{i,j}$ follows by switching
$x \leftrightarrow y$, $i \leftrightarrow j$, $ l_1\leftrightarrow l_2$,
$a_1 \leftrightarrow a_2$
and $K_1 \leftrightarrow K_2$.

{\it Linear version.} ---
In order to compare with the OFC model, we expand $\vec{F}_{i,j}$
to first order in $x$'s and $y$'s.
Specifying $\vec{f}_{L} = (-K_L x_{i,j},0)$ as in \cite{Brown,OFC},
one can easily show that a slip
of the block at $(i,j)$ results in
the following force redistribution\cite{note1,note2}:
\begin{eqnarray}
F^x_{i\pm 1,j} & \rightarrow & F^x_{i\pm 1,j}+\alpha_1 F^x_{i,j},\nonumber\\
F^x_{i,j\pm 1} & \rightarrow & F^x_{i,j\pm 1}
+ S_2 \sigma \alpha_1 F^x_{i,j}, \\
\label{redist}
F^y_{i\pm 1,j} & \rightarrow & F^y_{i\pm 1,j}
+ {S_1 \over \sigma} \alpha_2 F^y_{i,j}, \nonumber \\
F^y_{i,j\pm 1} & \rightarrow & F^y_{i,j\pm 1} +  \alpha_2 F^y_{i,j},\nonumber
\\
\vec{F}_{i,j} & \rightarrow & 0, \nonumber
\end{eqnarray}
where $S_1 \equiv (a_1 - l_1)/ a_1$
is the internal strain in the $x$ direction,
and similarly for $S_2$.
$\sigma \equiv K_2 /K_1$ and $\kappa \equiv K_L /K_1$
are measures of anisotropies in the couplings.
In the bulk, $\alpha_1$ and $\alpha_2$ are given by:
\begin{eqnarray}
\alpha_1 = {1 \over 2(1 + S_2 \sigma) + \kappa}  &;&
\alpha_2 = {1 \over 2(1 + S_1 / \sigma) }.
\label{alphas}
\end{eqnarray}
Their values at boundary sites depend on the boundary conditions (see below).

Using Eq.~(5),
our model can be described as a coupled map lattice
as was done in \cite{OFC}.
If we impose $y_{i,j} \equiv 0$ at all sites
as in \cite{Brown,OFC}, so that
(to first order) all $F^y_{i,j} = 0$,
we readily recover the OFC model in the limit
$S_2 \rightarrow 1$, i.e., for
{\em maximal\/} internal strain along the $y$ direction.
This also follows from the general relation
Eq.~(\ref{xcomponent}),
as $F^x$ becomes linear in $x$ for $S_2=1$ and $y=0$.
The physical rationale behind this correspondence is that
in \cite{Otsuka,Brown,OFC}, both the loading springs
and those in the array along $y$ were chosen as leaf springs, which
by definition cannot be extended along their length but
have a restoring force linear in the {\em transverse\/} displacement.
They act like fully stretched coil springs.
With leaf springs, the forces become scalar and the equations linear.
However, it introduces
an artificial asymmetry into the system, which to our knowledge
has no analog in nature. This manifests itself most clearly in the
elastic moduli\cite{elastic,note1}: For the OFC model and the like,
$C_{1111}=K_1 a_1/a_2$,
$C_{2222}=\infty$,
$C_{1122}=0$, and
$C_{1212}({\rm shear modulus})=K_2 a_2/a_1$.
Clearly, setting $K_1\equiv K_2$ does not render the model symmetric.
For our model, the asymmetry is absent:
$C_{1111}=K_1 a_1/a_2$,
$C_{2222}=K_2 a_2/a_1$,
$C_{1122}=0$, and
$C_{1212}=(K_1 K_2 S_1 S_2 a_1 a_2)/( K_1 S_1 a_1^2 + K_2 S_2 a_2^2)$.

For general strain ($0 \leq S_2 < 1$) and $F^y\equiv 0$,
the linear version of our model
coincides with the {\em anisotropic\/} OFC model\cite{OFC}
with our $S_2 \sigma$ corresponding to OFC's $K_2/K_1$.
Consequently, except in the unphysical limit $S_2\to 1$,
asymmetric loading on a symmetric
model ($\kappa = \sigma = 1, S_2 = S_1$)
gives rise to intrinsically asymmetric force redistributions.
The statement\cite{OFC} regarding the variance in the Gutenberg-Richter
law as a result of different elastic parameters has to
be reinterpreted, in the present context,
as a result of variances in both
internal strains and elastic parameters.

Allowing for $y_{i,j} \neq 0$,
Eqs.(5,\ref{alphas}) yield the {\em total\/}
change of force after a block {\em in the bulk\/} at $(i,j)$ slips by
a distance $(\delta x, \delta y)$:
\begin{eqnarray}
\delta F^x = -K_L \delta x = -\kappa \alpha_1 F^x_{i,j}  &  ;  &
\delta F^y = 0.
\label{deltaF}
\end{eqnarray}
The change on the boundary depends on the boundary conditions,
which are defined by
the number of neighbors a boundary block is attached,
{\em and\/} the way it is loaded.
Following OFC,
in ``free'' boundary conditions (FBC) the sites on edges have three
neighbors while those at corners have two.
In ``open'' boundary conditions (OBC)
all sites has four neighbors.
Both have {\em uniform\/} loading throughout the system by an amount
$\Delta F^x=K_L \Delta x$, where $\Delta x$ denotes the displacement
of the driving plate relative to the array.

For FBC,  the $\alpha$'s on the boundary
differ from the bulk values in Eq.~(\ref{alphas}):
\begin{eqnarray}
\begin{array}{lll}
\alpha_1^{(x=1,L)}={1 \over  1+2 S_2 \sigma + \kappa}, \mbox{~~}
& \alpha_1^{(y=1,L)}= {1 \over  2+  S_2 \sigma + \kappa}, \mbox{~~}
& \alpha_1^{\rm{(corner)}}= {1 \over  1+ S_2 \sigma + \kappa};
\\
\alpha_2^{(x=1,L)}={1 \over  2+  S_1/\sigma }, \mbox{~~}
& \alpha_2^{(y=1,L)}= {1 \over  1+ 2S_1/\sigma }, \mbox{~~}
& \alpha_2^{\rm{(corner)}}= {1 \over  1+ S_1/\sigma}.
\end{array}
\label{fbcalphas}
\end{eqnarray}
It turns out then that Eq.~(\ref{deltaF}) holds at all sites.
While the change of $F^x$ is balanced on the average by loading,
the spatial mean $\bar F^y$ is always zero due to Newton's third law and
the fact that the system is isolated for FBC.
Since a slip is decided by $\sqrt{(F^x)^2+(F^y)^2}-F_{th}$,  the
two components are locally coupled in the dynamics.
We find
in the steady state that the fluctuation of
$F^y$ tends to zero, so that this
coupling to a conservative field is irrelevant\cite{note3}, i.e.,
the OFC limit is recovered.

For OBC, Eq.~(\ref{alphas}) holds at all sites but
Eq.~(\ref{deltaF}) is modified to $\delta F^y/F^y <0$ at boundary
sites due to their coupling to the external imaginary blocks.
This implies that a system with arbitrary initial
spatial mean, $\bar F^y$, will always flow in steady state
to the $\bar F^y=0$
(stable) fixed point (i.e., the OFC model).
We conclude, for OBC too, that the steady state of
our linearized version is the same as that of the OFC model.

We now remark on the boundary conditions.
We first reveal a flaw in previous treatments of OBC\cite{OFC}.
Effectively, the ``imaginary layer of blocks'' in OBC corresponds
to a rigid frame that attaches to the array by springs.
Spatially uniform loading in Ref.~\cite{OFC} implies no relative
displacement between the frame and the array during loading.
Since the imaginary layer never slips, it is permanently pinned
on the rough surface.
Although this is an unphysical setup,  it has not been
appreciated before in \cite{OFC,grassberger} because it is not
obvious in the $F_{i,j}$ configurations.
Working with the linearly related $x_{i,j}$ instead,
we show in Fig.~1 how the initially square array of blocks is distorted
due to the pinning.
At long times, the distortion can be arbitrarily large so that
there is no meaningful steady state for this model.
To make physical sense,  the frame has to move along
with the driving plate.
It is then intuitively clear that loading cannot be
uniform: due to pulling and pushing by the frame,
boundary blocks are loaded more than in the bulk
(e.g., $\Delta F^x_{1,j}=(1+\kappa^{-1})\Delta F^x_{\rm{bulk}}$).
Incorporating such non-uniformity,  we observe numerically\cite{note1}
that the model becomes noncritical and reaches periodic states much like
with periodic boundary conditions\cite{grassberger}.
Second, for FBC, we implicitly assume in
Eqs.~(5) and (\ref{fbcalphas})
that the internal strains are somehow sustained. Otherwise,
due to missing neighbors at the boundary,
the equilibrium positions of blocks would be shifted inward relative to
a regular lattice.
In practice, the locations of loading springs on the driving plate
may be adjusted to eliminate the shifts.
In any case, bulk properties are not affected because
the shifts decay exponentially into the bulk
over a distance[=$1/\ln(2+\kappa)$] of order unity\cite{note1}.
The use of leaf springs along $y$ direction
in \cite{Otsuka,Brown,OFC} may also be
interpreted as another means of imposing the strains.

{\it Nonlinear version.} ---
Eq.~(\ref{xcomponent}) allows for an investigation
of nonlinear effects which naturally arise from the full forcing.
However, it is no longer possible to describe the model
as a coupled map lattice [cf. Eq.~(5)].
Instead, one has to keep track of the displacements in order
to determine whether $|\vec{F}|>F_{th}$ ---
if so, one has to solve
for the displacements ($\tilde{x}, \tilde{y}$)
that defines the zero-force position of the block
in terms of its environment.
To $n^{\rm th}$ order, this means to solve two coupled equations
of the form [for brievity, $q\equiv (i,j)$]:
\begin{eqnarray}
\tilde F^x_q &\equiv & 0 = f_L^x -\sum_{p}  K_{p,q} \{ u_{p,q} - \sum_{m=0}^n
  {1 \over m!} [ u_{p,q} {\partial \over \partial x}
  + v_{p,q} {\partial \over \partial y} ]^m
  G_{p,q}(x,y)|_{(x,y)=(0,0)} \},
\label{expansion}
\end{eqnarray}
plus a corresponding equation ($\tilde F^y_q\equiv 0$) obtained by the
transformation after Eq.~(\ref{xcomponent}).
Here $p$ denotes the four nearest neighbors $(i\pm 1, j\pm 1)$,
and the symbols stand for:
\begin{eqnarray}
\begin{array}{lll}
K_{p,q} =K_1,\mbox{~~~}
& G_{p,q} (x,y) ={(x\mp a_1) l_1 \over
   \sqrt{(x \mp a_1)^2 + y^2} },
& \mbox{~~~for $p=(i\pm 1,j)$}; \\
K_{p,q}  =K_2,\mbox{~~~}
& G_{p,q} (x,y) ={ x l_2 \over \sqrt{(y \mp a_1)^2 + x^2} };
& \mbox{~~~for $p=(i,j\pm 1)$}; \\
u_{p,q}  =\tilde{x_q}-x_p,\mbox{~~~}
& v_{p,q} =\tilde{y_q}-y_p,\mbox{~~~}
& \mbox{~~~for $p=(i\pm 1,j),(i,j\pm 1)$}.
\end{array}
\label{symbols}
\end{eqnarray}
Note that we have implicitly assumed that no
nearest neighbors can be supercritical at the same time, due to
infinitesimal loading rate.

Apparently this program is not computationally efficient for large $n$.
But $n=2$ is simple: It is easy to see that
all second order terms ($\tilde{x}^2, \tilde{y}^2,
\tilde{x}\tilde{y}$) cancel {\em in the bulk\/}
(i.e., when $(i,j)$ has four nearest neighbors),
resulting in two coupled {\em first\/}
order equations for $(\tilde{x},\tilde{y})$ in terms of
$F$ and $(x,y)$ before slipping:
\begin{eqnarray}
{F^x_{i,j} \over K_1}
& = & (\tilde x_{i,j}-{x}_{i,j})[\alpha_1^{-1}
 + {\sigma l_2\over a_2^2} (y_{i,j+1} - y_{i,j-1})] \nonumber \\
& & + (\tilde y_{i,j}-{y}_{i,j})[{l_1\over a_1^2} (y_{i+1,j}-y_{i-1,j})
 + {\sigma l_2 \over a_2^2}(x_{i,j+1} - x_{i,j-1})],
\label{2ndorder}
\end{eqnarray}
plus a corresponding equation for $F^y_{i,j}$.
Note that it yields nonlinear dependence
of $(\tilde x,\tilde y)$ on $(x,y)$ {\em via\/} those terms proportional to
$l/a^2$.

Based on Eq.~(\ref{2ndorder}), we have performed simulations
for different boundary conditions. Due to missing neighbors of the
blocks on the edges for FBC, second order terms in $(\tilde x,\tilde y)$
survive and one has to solve higher
order ($3^{\rm rd}$ and $4^{\rm th}$)
equations for the equilibrium positions of
the boundary blocks. Remarkably, even with substantial nonlinearities
(determined by $l/a^2$), in none of the cases of FBC, OBC and
PBC do we find the
critical behavior to deviate from that of the linear cases.

To summarize, we have investigated the critical behavior of
a general 2D spring-block model of earthquakes,
within the context of the BK model under the quasi-static limit.
We discover that internal strains are important ingredients in
the variance of the Gutenberg-Richter law.
Nonlinear perturbations, to which the OFC model displays remarkable stability,
are not introduced arbitrarily
but generated generically
at the microscopic level from the elastic spring-block interactions.
Among possible extensions,
the case of internal compressional strains
(i.e., $S<0$),
and of different loadings, including those along
both directions [i.e., $f_L=-( K_L^x x, K_L^y y)$] and spatially non-uniform
ones (e.g., shear applied through the boundary),
are being pursued\cite{note1}.

J.V.A wishes to acknowledge support from the Natural Sciences and Engineering
Research Council of Canada and
the Danish Natural Science Research Council under Grant No. 11-9863-1.
J.V.A. and J.M. thank Y. Brechet, D. Sornette and M. Zucherman
for useful discussions.
K.-t.L thanks J.-H. Wang for an illuminating discussion,
and the National Science Council of ROC for support.

\newpage
\begin{center}{\bf Figure captions.}\end{center}

\noindent {\bf Fig.~1.}
Snapshot of a configuration of blocks after 2000 avalanches, for $L=20$,
$a_1=a_2=1$, $l_1=l_2=0.1$ and $\sigma=\kappa=K_1=F_{th}=1$,
showing the unphysical effect of pinned frame (denoted by $\Box$) in OBC.
The array is pulled to the right.
The system evolves according to Eq.~(\ref{2ndorder}).



\newpage
\begin{thebibliography}{99}
\bibitem{Burridge} R. Burridge and L. Knopoff,
Bull. Seis. Soc. Am. {\bf 57}, 341 (1967).

\bibitem{Langer}
J. M. Carlson and J. S. Langer, Phys. Rev. Lett. {\bf 62}, 2632 (1989);
Phys. Rev. A {\bf 40}, 6470 (1989).

\bibitem{Nakanishi}
H. Nakanishi, Phys. Rev. A {\bf 41}, 7086 (1990);
Phys. Rev. A {\bf 43}, 6613 (1991).

\bibitem{Otsuka}
M. Otsuka, J. Phys. Earth {\bf 20}, 34 (1972);
Phys. Earth Planet. Inter. {\bf 6}, 311 (1972).

\bibitem{Brown}
S. R. Brown, C. H. Scholz, and J. B. Rundle,
Geophys. Res. Lett. {\bf18}, 215 (1991).

\bibitem{Bak} P. Bak, C. Tang, and K. Wiesenfeld, Phys. Rev. Lett.,
{\bf 59}, 318 (1987).

\bibitem{OFC} Z. Olami, J. S. Feder, and K. Christensen,
Phys. Rev. Lett. {\bf 68}, 1244 (1992);
K. Christensen and Z. Olami, Phys. Rev. A {\bf 45}, 665 (1992).

\bibitem{Xu} H.-J. Xu, B. Bergersen and K. Chen, J. Phys. A: Math. Gen.
{\bf 25}, L1251 (1992);
H.-J. Xu, B. Bergersen and K. Chen,
preprint (1994).

\bibitem{grassberger}
J. E. S. Socolar, G. Grinstein and C. Jayaprakash, Phys. Rev. E {\bf 47},
2366 (1993);
P. Grassberger, Phys. Rev. E., {\bf 49}, 2436 (1994).

\bibitem{Langer1}
J. M. Carlson, J. S. Langer and B. E. Shaw, Rev. Mod.
Phys. {\bf 66}, 657 (1994).

\bibitem{Joergen}
J. V. Andersen, Y. Brechet and H. J. Jensen,
Europhys. Lett. {\bf 26}, 13 (1994);
J. V. Andersen, Phys. Rev. B {\bf 49}, 9981 (1994).

\bibitem{Gutenberg}
B. Gutenberg and C. F. Richter, Ann. Geophys. {\bf 9}, 1 (1956).

\bibitem{Pacheco}
J. F. Pacheco, C. H. Scholz, and L. R. Sykes,
Nature {\bf 355}, 71 (1992).

\bibitem{elastic}
See, e.g., L.D. Landau and E.M. Lifshitz, Theory of Elasticity, 3nd ed.
(Pergamon Press, Oxford, 1986)

\bibitem{note1}
K.-t. Leung, J. M\"{u}ller and J. V. Andersen (unpublished).

\bibitem{note2}
In 2D, the equations of motion in the continuum approach of
Ref.~\cite{Langer,Langer1}
to first order in the displacements $\vec{U}\equiv (U_x,U_y)$,
including strain $S$, are similarly modified:
\begin{eqnarray}
\partial_t^2 U_x & =& \xi^2(\partial_x^2 U_x + S \partial_y^2 U_x) - U_x
- \partial_t U_x \phi(|\partial_t \vec{U}|) \nonumber \\
\partial_t^2 U_y & =& \xi^2(S\partial_x^2 U_y +  \partial_y^2 U_y)
- \partial_t U_y \phi(|\partial_t \vec{U}|), \nonumber
\end{eqnarray}
where $\phi(|\partial_t \vec{U}|)$ is a speed-dependent friction
term and $\xi$ a characteristic length.
The Eq.~(9) of Ref.~\cite{Langer1} may be obtained by setting
$S=1$ and $U_y=0$.

\bibitem{note3}
In analogy to critical dynamics
[see, e.g., P.C. Hohenberg and B.I. Halperin, Rev. Mod. Phys.
{\bf 49}, 435 (1977)],
we may consider the model as a
generalized coupled map lattice
with an auxillary field $F^y$ conserved at $\bar F^y\neq 0$.
We found that the value of $\bar F^y$ is relevant,
e.g., the exponent $B$ in Gutenberg-Richter law
decreases as $\bar F^y$ increases,
implying that $F^y\equiv 0$ is special.

\end{thebibliography}
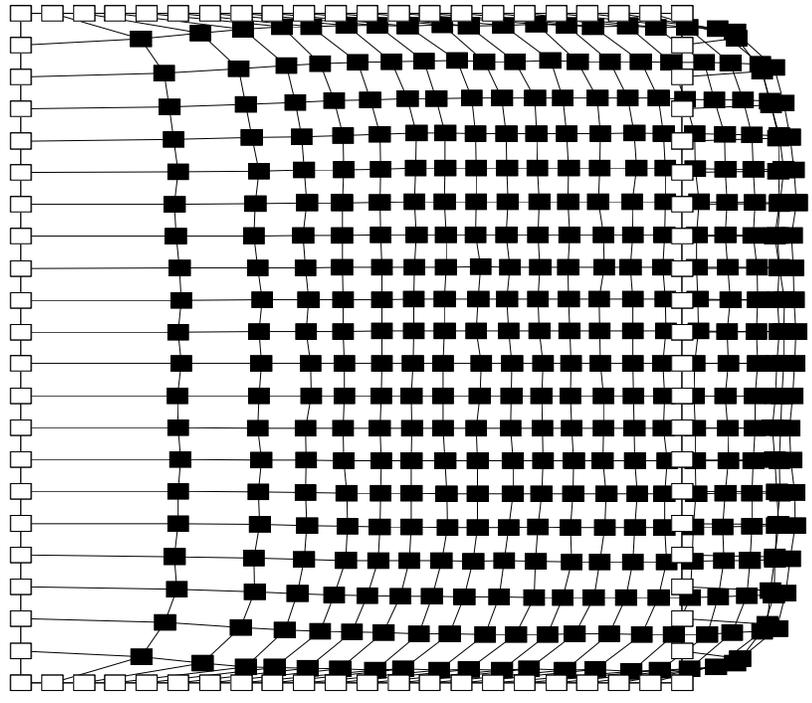
\end{document}